\documentclass{bmcart}

\RequirePackage{hyperref}
\usepackage[utf8]{inputenc} %unicode support
\usepackage{graphicx}
\graphicspath{{Graphics/}{./Graphics/}}

\startlocaldefs
\endlocaldefs

\begin{document}

%%% Start of article front matter
\begin{frontmatter}

\begin{fmbox}
\dochead{Research}

%%%%%%%%%%%%%%%%%%%%%%%%%%%%%%%%%%%%%%%%%%%%%%
%%                                          %%
%% Enter the title of your article here     %%
%%                                          %%
%%%%%%%%%%%%%%%%%%%%%%%%%%%%%%%%%%%%%%%%%%%%%%

\title{Engineering Structural Robustness in Power Grid Networks Susceptible to Community Desynchronization}

%%%%%%%%%%%%%%%%%%%%%%%%%%%%%%%%%%%%%%%%%%%%%%
%%                                          %%
%% Enter the authors here                   %%
%%                                          %%
%% Specify information, if available,       %%
%% in the form:                             %%
%%   <key>={<id1>,<id2>}                    %%
%%   <key>=                                 %%
%% Comment or delete the keys which are     %%
%% not used. Repeat \author command as much %%
%% as required.                             %%
%%                                          %%
%%%%%%%%%%%%%%%%%%%%%%%%%%%%%%%%%%%%%%%%%%%%%%

\author[
   addressref={addr_uw},                   % id's of addresses, e.g. {aff1,aff2}
%   corref={aff1},                       % id of corresponding address, if any
%   noteref={n1},                        % id's of article notes, if any
   email={dylewsky@uw.edu}   % email address
]{\inits{D}\fnm{Daniel} \snm{Dylewsky}}
\author[
   addressref={addr_pnnl},
   email={xiu.yang@pnnl.gov}
]{\inits{X}\fnm{Xiu} \snm{Yang}}
\author[
   addressref={addr_pnnl},
   email={alexandre.tartakovsky@pnnl.gov}
]{\inits{A}\fnm{Alexandre} \snm{Tartakovsky}}
\author[
   addressref={addr_uw},
   email={kutz@uw.edu}
]{\inits{JN}\fnm{J. Nathan} \snm{Kutz}}

%%%%%%%%%%%%%%%%%%%%%%%%%%%%%%%%%%%%%%%%%%%%%%
%%                                          %%
%% Enter the authors' addresses here        %%
%%                                          %%
%% Repeat \address commands as much as      %%
%% required.                                %%
%%                                          %%
%%%%%%%%%%%%%%%%%%%%%%%%%%%%%%%%%%%%%%%%%%%%%%

\address[id=addr_uw]{%                           % unique id
  \orgname{Department of Applied Mathematics, University of Washington}, % university, etc
%  \street{Waterloo Road},                     %
  \postcode{98195-2420}                                % post or zip code
  \city{Seattle, WA},                              % city
  \cny{US}                                    % country
}
\address[id=addr_pnnl]{%
  \orgname{Pacific Northwest National Laboratory},
  \street{902 Battelle Blvd, Richland, WA 99354},
  \postcode{99354}
  \city{Richland},
  \cny{Washington}
}

%%%%%%%%%%%%%%%%%%%%%%%%%%%%%%%%%%%%%%%%%%%%%%
%%                                          %%
%% Enter short notes here                   %%
%%                                          %%
%% Short notes will be after addresses      %%
%% on first page.                           %%
%%                                          %%
%%%%%%%%%%%%%%%%%%%%%%%%%%%%%%%%%%%%%%%%%%%%%%

\begin{artnotes}
%\note{Sample of title note}     % note to the article
\note[id=n1]{Equal contributor} % note, connected to author
\end{artnotes}

\end{fmbox}% comment this for two column layout

%%%%%%%%%%%%%%%%%%%%%%%%%%%%%%%%%%%%%%%%%%%%%%
%%                                          %%
%% The Abstract begins here                 %%
%%                                          %%
%% Please refer to the Instructions for     %%
%% authors on http://www.biomedcentral.com  %%
%% and include the section headings         %%
%% accordingly for your article type.       %%
%%                                          %%
%%%%%%%%%%%%%%%%%%%%%%%%%%%%%%%%%%%%%%%%%%%%%%

\begin{abstractbox}

\begin{abstract} % abstract
Networked power grid systems are susceptible to a phenomenon known as Coherent Swing Instability (CSI), in which a subset of machines in the grid lose synchrony with the rest of the network.  We develop network level evaluation metrics to (i) identify community substructures in the power grid network, (ii) determine weak points in the network that are particularly sensitive to CSI,  and (iii) produce an engineering approach for the addition of transmission lines to reduce the incidences of CSI in existing networks, or design new power grid networks that are robust to CSI by their network design.  For simulations on a reduced model for the American Northeast power grid, where a block of buses representing the New England region exhibit a strong propensity for CSI, we show that modifying the network’s connectivity structure can markedly improve the grid’s resilience to CSI.  Our analysis provides a versatile diagnostic tool for evaluating the efficacy of adding lines to a power grid which is known to be prone to CSI. This is a particularly relevant problem in large-scale power systems, where improving stability and robustness to interruptions by increasing overall network connectivity is not feasible due to financial and infrastructural constraints. 
\end{abstract}

%%%%%%%%%%%%%%%%%%%%%%%%%%%%%%%%%%%%%%%%%%%%%%
%%                                          %%
%% The keywords begin here                  %%
%%                                          %%
%% Put each keyword in separate \kwd{}.     %%
%%                                          %%
%%%%%%%%%%%%%%%%%%%%%%%%%%%%%%%%%%%%%%%%%%%%%%


% MSC classifications codes, if any
%\begin{keyword}[class=AMS]
%\kwd[Primary ]{}
%\kwd{}
%\kwd[; secondary ]{}
%\end{keyword}

\end{abstractbox}
%
%\end{fmbox}% uncomment this for twcolumn layout

\end{frontmatter}

%%%%%%%%%%%%%%%%%%%%%%%%%%%%%%%%%%%%%%%%%%%%%%
%%                                          %%
%% The Main Body begins here                %%
%%                                          %%
%% Please refer to the instructions for     %%
%% authors on:                              %%
%% http://www.biomedcentral.com/info/authors%%
%% and include the section headings         %%
%% accordingly for your article type.       %%
%%                                          %%
%% See the Results and Discussion section   %%
%% for details on how to create sub-sections%%
%%                                          %%
%% use \cite{...} to cite references        %%
%%  \cite{koon} and                         %%
%%  \cite{oreg,khar,zvai,xjon,schn,pond}    %%
%%  \nocite{smith,marg,hunn,advi,koha,mouse}%%
%%                                          %%
%%%%%%%%%%%%%%%%%%%%%%%%%%%%%%%%%%%%%%%%%%%%%%

%%%%%%%%%%%%%%%%%%%%%%%%% start of article main body
% <put your article body there>

%%%%%%%%%%%%%%%%
%% Background %%
%%

%	INTRODUCTION
\section*{Introduction}

Disruptions of power grid systems can have a severe, negative impact on performance and lead to  
{\em Coherent Swing Instability} (CSI)~\cite{susuki2011nonlinear,susuki2011coherent,susuki2012nonlinear}, whereby a subset of machines in the grid lose synchrony with the rest of the network, thus shutting the entire network down and leading to unacceptable blackouts.
CSI is in essence a manifestation of community structure in a networked system, a collective dynamical divergence of one subgroup of nodal oscillations from another. Instigated by the work of Girvan and Newman on modular structures in networks \cite{Girvan2002,Newman2004,Newman2006}, a great deal of attention has been devoted to the development of methods for identification and characterization of modular components. {\em Community detection} has become a broadly-defined term which is used to refer to a variety of such approaches. A comprehensive review of its uses is presented by Schaub et. al. \cite{Schaub2017}. Much of the prominent work in the field has focused on topological approaches, i.e. methods which take as input an adjacency matrix describing a (weighted or unweighted, directed or undirected) graph. As Schaub notes, however, dynamics on a network are constrained by topology but cannot be fully described by it.

This distinction has led to work on community detection methods which place primary importance on dynamics. However, as many networks of interest lack a known set of laws governing their evolution (e.g. neurological, social, transportation, epidemiological, etc.), much of this research has modeled dynamics with Markovian diffusion processes describing random flows on the network \cite{Delvenne2010,Rosvall2014,Bacik2016}. Power grids have an advantage over these systems in that the physics of generator oscillations and current transmission are well-understood: taken in isolation, each node of the network behaves predictably. Only when they are combined on a complex and non-symmetric graph structure do they begin to exhibit the group behaviors which resist simple characterization. In this work we seek to capitalize on this property by applying a dynamics-focused community detection perspective to simulation data generated by a realistic, machine-level power grid model.

Previous work on applying network-topological analysis to real-world power grids has met with mixed success. The extensive theoretical framework that has been developed in the field of Complex Networks has given rise to a variety of methods which assess functional properties of power grids directly from their network topology, using metrics such as node centrality, betweenness, degree distribution, community structure, and clustering coefficients \cite{cuadra15,albert04,bompard15,sole08,zhou15,xu14,pagani14,pahwa13}. However, topological methods alone have consistently failed to fully account for observed network vulnerabilities \cite{bompard09,bompard10,pagani13,hines10}. This has led to a variety of hybrid approaches which incorporate electrical properties of the system not captured by its graph structure, often by using them to assign edge weights or to compute modified versions of existing topological metrics \cite{nasiruzzaman11,hines10,yan13,bompard10,bompard12}. The goal of these studies has generally been to develop a heuristic to estimate a power grid's vulnerability to failure, which they then validate using historical data or numerical simulations. In this work we present an alternative method which evaluates the functional consequences of structural modifications directly from simulation data. While this approach has previously been applied to steady-state systems \cite{ieee09}, its use for grid disturbances has generally been dismissed as infeasible due to the combinatorially large search space of possible faults and network structures \cite{hines10}. Our contribution is a methodology for incorporating topological properties (specifically, community structure) and statistics on measurement data to significantly reduce this space. Our technique identifies a set of candidate locations for single-line additions to an existing network; this is a small structural perturbation relative to the size of the full grid and therefore unlikely to significantly change its global topological characteristics, such as degree distribution. Robustification occurs not by directly tuning some topological (or hybrid topological/electrical) metric, but rather by using such a metric to inform a minor structural modification. This lends itself to practical engineering application, as it suggests an inexpensive change that could be made to an existing grid rather than a design principle for the construction of a new grid from the ground up.

Our data-driven approach is motivated by the proliferation of real-time monitoring strategies for power grids that have been deployed in recent years \cite{messina2015wide,kezunovic2013role,barocio2013detection}, with event location strategies gaining increasing attention in order to localize pernicious effects~\cite{li2010online,mei2008clustering,bhui2016application}. These strategies are directed to provide a system-wide awareness of events such as faults and other disturbances, taking advantage of the increasing coverage of {\em wide area measurement systems} (WAMS) technology which enables the implementation of wide area emergency and restorative control applications~\cite{li2010online,mei2008clustering,bhui2016application}.
Even though these strategies have achieved positive results, additional technical challenges arise as the modern WAMS-generated data become high dimensional and more distributed thorough large areas of the system. 
We propose an additional technique for power grid network robustification.  Specifically, we show that the robustness of the network can be diagnosed from ensemble fault simulations. Moreover, the power grid can be made significantly more robust to disturbances with proper engineering of the network design and attention to the community structure observed in instances of CSI. Such considerations are critical in considering future power systems deployments, or for upgrading current networks in order to circumvent susceptibility to CSI.

In this paper, we detail a simulation model used for characterizing the power grid dynamics and disruptions.  We further introduce a procedure for dynamics-based community detection based on the results of these simulations, showing through diagnostic tools that the network's sensitivity to CSI depends on the location and severity of a fault.  This allows us to provide an engineering approach capable of characterizing network connectivity modifications capable of robustifying the network to decoherence.

%	NUMERICAL SIMULATIONS OF POWER GRIDS
\section*{Numerical Simulations of Power Grids}
\label{sec:pg}

\begin{figure}[h!]
	\includegraphics[width=0.9\columnwidth]{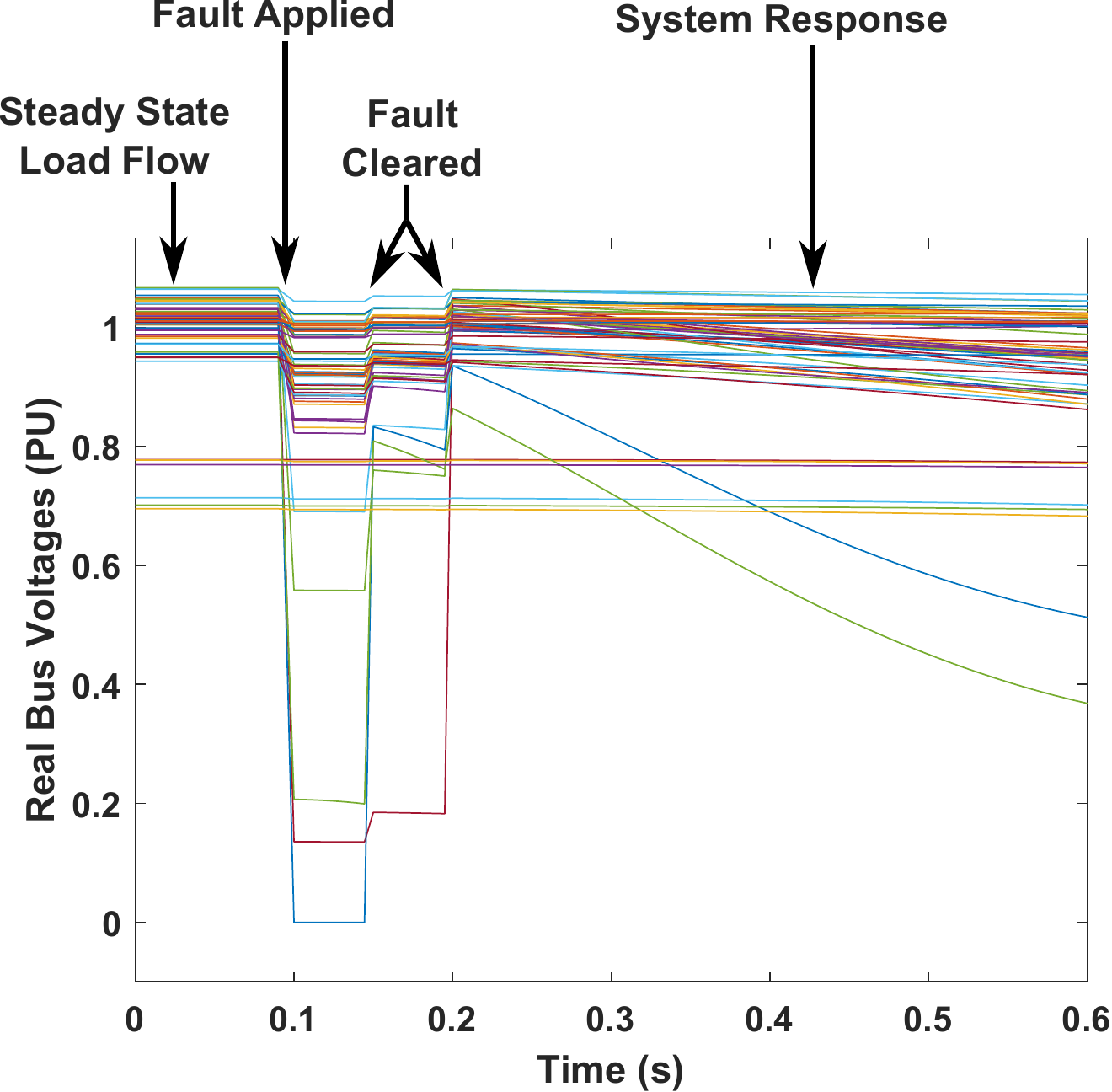}
	\caption{\csentence{Power System Toolbox: Sample Simulation}
	Stages of a grid simulation in PST: The system is in a steady state until a fault is applied at 0.1s. The fault is cleared in two stages, at 0.15 and 0.2s. After the fault has been cleared, the system has been perturbed from its steady state so it continues to evolve dynamically}
	\label{fig:PST_demo}
\end{figure}

In this section, we highlight the numerical simulation architecture used to evaluate power grid systems and their connectivity structure.  Importantly, a prescription of the disturbances applied to the network to induce CSI is considered in order to evaluate the network robustness in a principled way.

\subsection*{The Power System Toolbox}
Power grid simulations for this study were produced using Power System Toolbox (PST), a Matlab software package originally developed by Kwok W. Cheung and Joe Chow of Rensselaer Polytechnic Institute~\cite{chow92}. Supplied with both the topological structure of a power network and the specific electromechanical parameters of the grid's generators, nodes, and lines, PST performs dynamic simulations of both steady-state and nonequilibrium dynamics. In this study we employ system fault simulations structured as follows:

\begin{enumerate}

\item The system is initialized in a steady state for power flow through the network by solving the nonlinear algebraic network equations which specify the load flow problem.

\item The dynamic portion of the simulation is initiated by applying a transient three-phase fault to a single line in the network, as though the line were brought into momentary contact with a grounding object such as a tree. 

\item When a fault occurs, power system protection equipment acts to isolate the disturbance. If the fault is transient, the line can be reconnected after a short time. PST treats this as a two-step process, clearing the fault first at the near end and then at the remote end of the line. These two time intervals, labeled in this paper as $\tau_1$ and $\tau_2$, are supplied as input parameters in simulations.  Figure~\ref{fig:PST_demo} illustrates the two time scales, $\tau_1$ and $\tau_2$, for a simulation of the northeastern power grid system.  Our interest is in the dynamics following the fault application and fault clearing time scales.

\item When the fault is cleared, the grid recovers its full original network structure. The fault has perturbed it from its initial steady-state configuration, so dynamic evolution continues. We continue the simulation for a long time (relative to the fault duration) and analyze the network's response to the disturbance.  The CSI often is induced by a disturbance event (fault application) for which the power grid system does not recover to its original stable (steady-state) behavior.

\end{enumerate}

\begin{figure}[h!]
	\includegraphics[width=0.9\columnwidth]{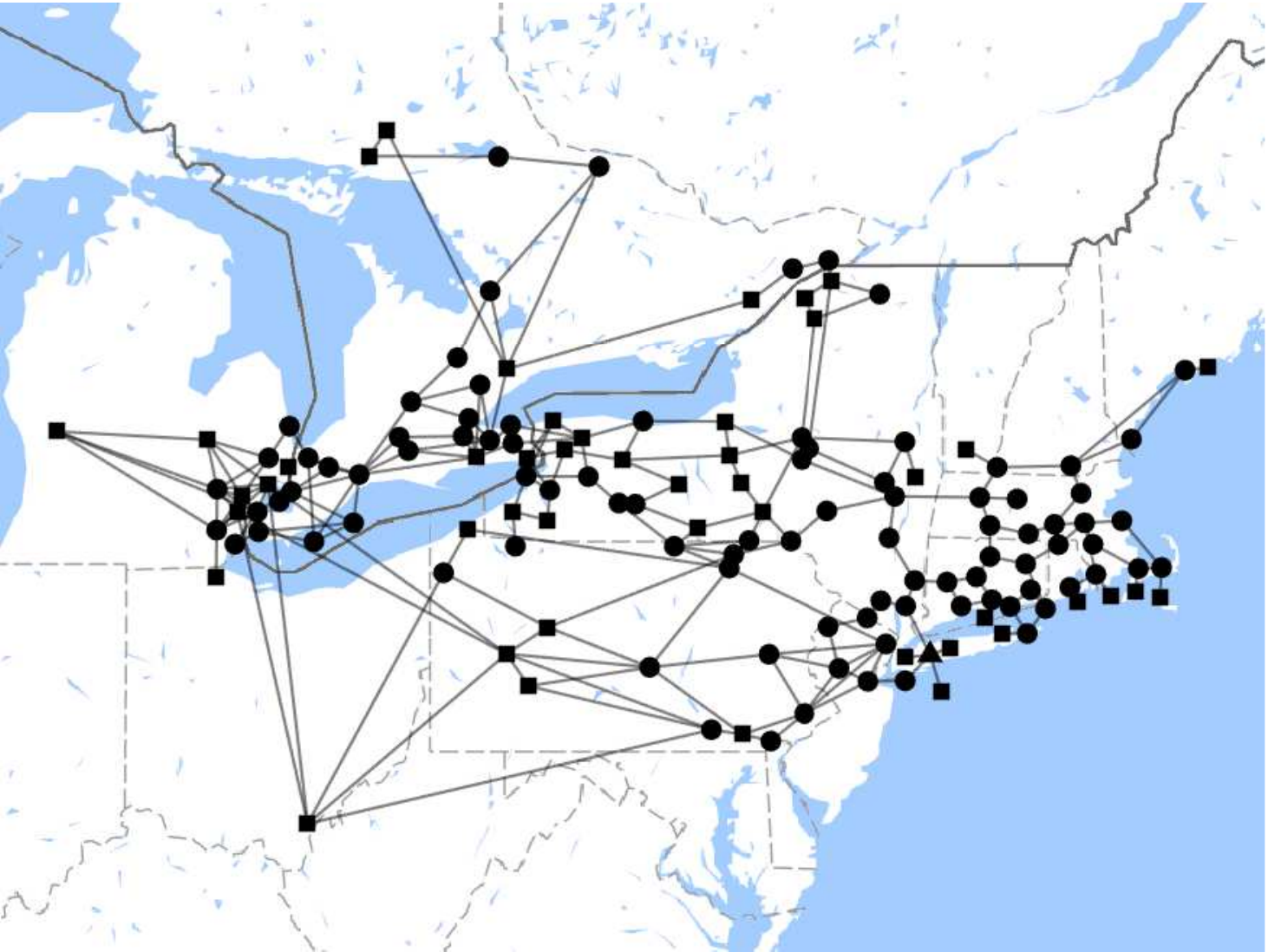}
	\caption{\csentence{NPCC System Map}
	The NPCC 140-Bus system is a reduced model for the power grid of the northeastern United States and Canada, plotted here overlaid on a map of the region. In this figure only, bus types are distinguished by the nodes' symbols (circles are load buses, squares are generator buses, and triangles are swing buses).}
	\label{fig:NPCC_map}
\end{figure}

\subsection*{The NPCC 140-Bus System}
Simulations in this study are carried out on the NPCC 140-bus test system, which is a reduced model based on the power grid of the North American northeast (Fig. \ref{fig:NPCC_map}). This network was chosen because its machine parameters are representative of those in a major real-world grid and its graph structure is sufficiently large and complex that it gives rise to coherent dynamics at subnetwork level.

\section*{Discovering Community Structure in the Grid}
\label{sec:community}
Simulations of the northeastern power grid are sufficient to illustrate many of the key features of power grid networks and their induced CSI.  By perturbing the various nodes of this specific network, we can characterize the instability structures, and their commonalities, induced in the power grid dynamics.

\begin{figure}[h!]
	\includegraphics[width=0.9\columnwidth]{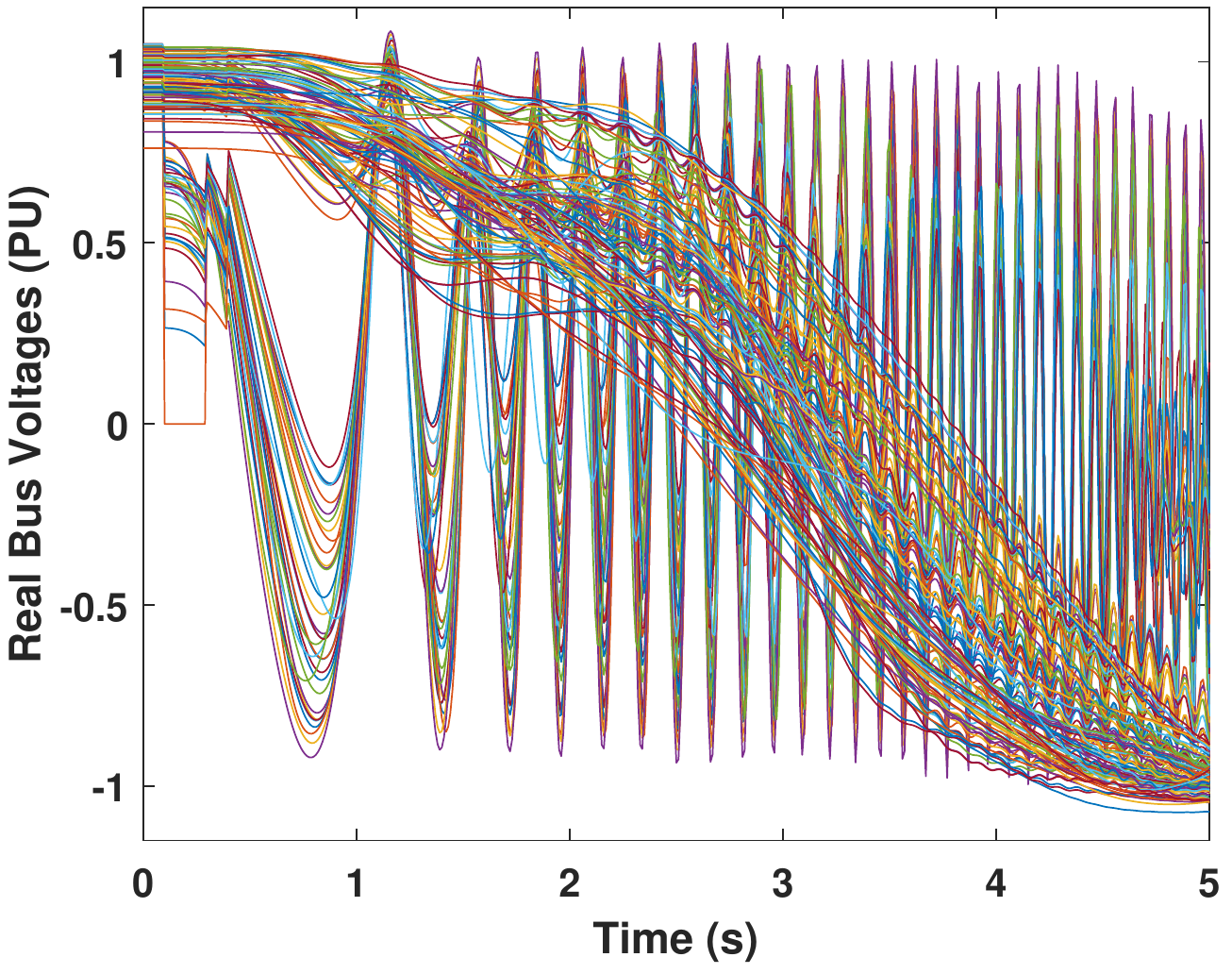}
	\caption{\csentence{Coherent Swing Instability}
	CSI in numerical simulations: Plotting bus voltages shows two qualitatively different types of dynamics. Each bus belongs to one of two coherent groups, one of which varies at a slow, consistent time scale and the other of which oscillates with linearly increasing frequency.}
	\label{fig:unst_voltages}	
\end{figure}

\subsection*{Identifying Coherent Swing Instabilities}
The focus of this specific study is the phenomenon of CSI, in which a subgroup of buses which are strongly coupled to one another, but only weakly coupled to other nodes, collectively lose synchronicity with the remainder of the network. Real-world power systems implement controllers to damp the oscillations of relative rotor angles which give rise to CSI, but no such safeguards are implemented in the PST simulation toolbox. The onset of CSI therefore is manifested as a clear qualitative transition in the dynamics of a subset of the buses. Specifically, a group of machines will begin to oscillate with linearly increasing frequency while the remainder of the network continues to evolve with dynamics on a slower and roughly constant time scale (Fig. \ref{fig:unst_voltages}).  Although the PST model does not accurately model the behavior of the unstable network, since local controllers would activate to dampen growing oscillations, it does highlight the lack of robustness of the network to the intrinsic dynamics induced by the disturbances.  By engineering a more robust system, the intrinsic dynamics itself acts to stabilize the system.   This aspect of engineering a power grid network is considered below in the section entitled \textit{Engineering Network Structure to Reduce CSI}.

\begin{figure}[h!]
	\includegraphics[width=0.9\columnwidth]{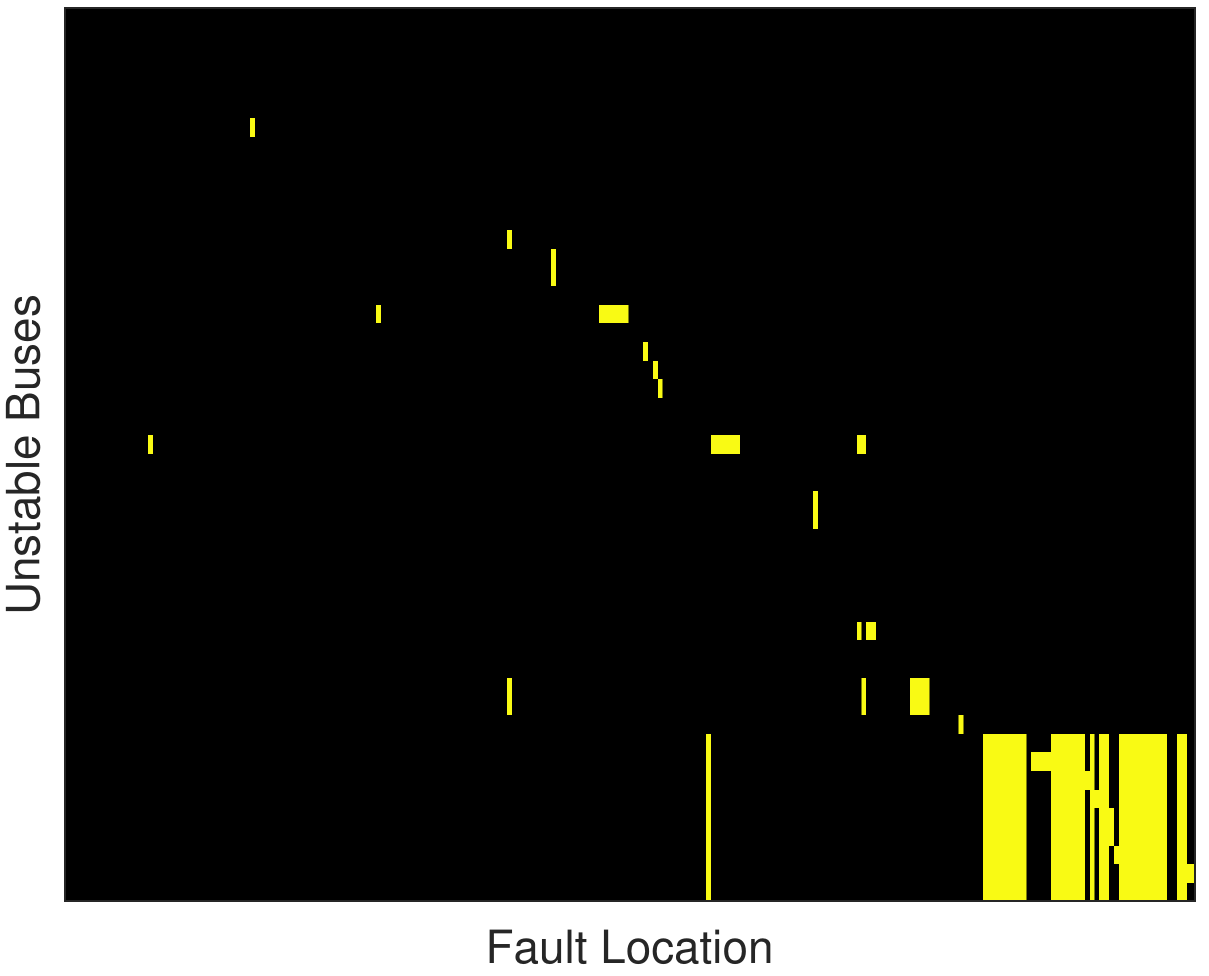}
	\caption{\csentence{Unstable Buses Based on Fault Location}
	A systematic exploration of the network locations of unstable dynamics as a function of the line where the initial fault was applied. Each column represents a different simulation with a different fault location, with yellow pixels denoting which buses exhibited instability during that run.}
	\label{fig:unstable_bloc}	
\end{figure}

\subsection*{Community Detection}
We begin by investigating the incidence of CSI by systematically applying faults to each line of the network in succession and tracking which buses (if any) exhibit unstable dynamics as a result. The results, plotted in Fig.~\ref{fig:unstable_bloc}, suggest that the majority of instabilities take place in a particular subgroup of buses. To formally characterize this structure we approach it as a network community detection problem: by treating the buses and fault locations as two disjoint populations of nodes with (unweighted, undirected) connections given by the nonzero elements in Fig. \ref{fig:unstable_bloc}, we build a square adjacency matrix which casts the results as a bipartite instability matrix (with $(N_{\rm{bus}} + N_{\rm{line}})$ rows and columns).

In this form, the results are amenable to any network-topological community detection scheme desired. Results presented in this paper use the Adaptive BRIM algorithm, developed by Michael J. Barber for modularity-based community detection in bipartite networks \cite{Barber2007}. Very similar results were obtained using other bipartite algorithms, including those introduced in Newman (2008) \cite{Newman2006} and Liu and Murata (2009) \cite{Liu2009}.

\begin{figure}[h!]
	\includegraphics[width=0.9\columnwidth]{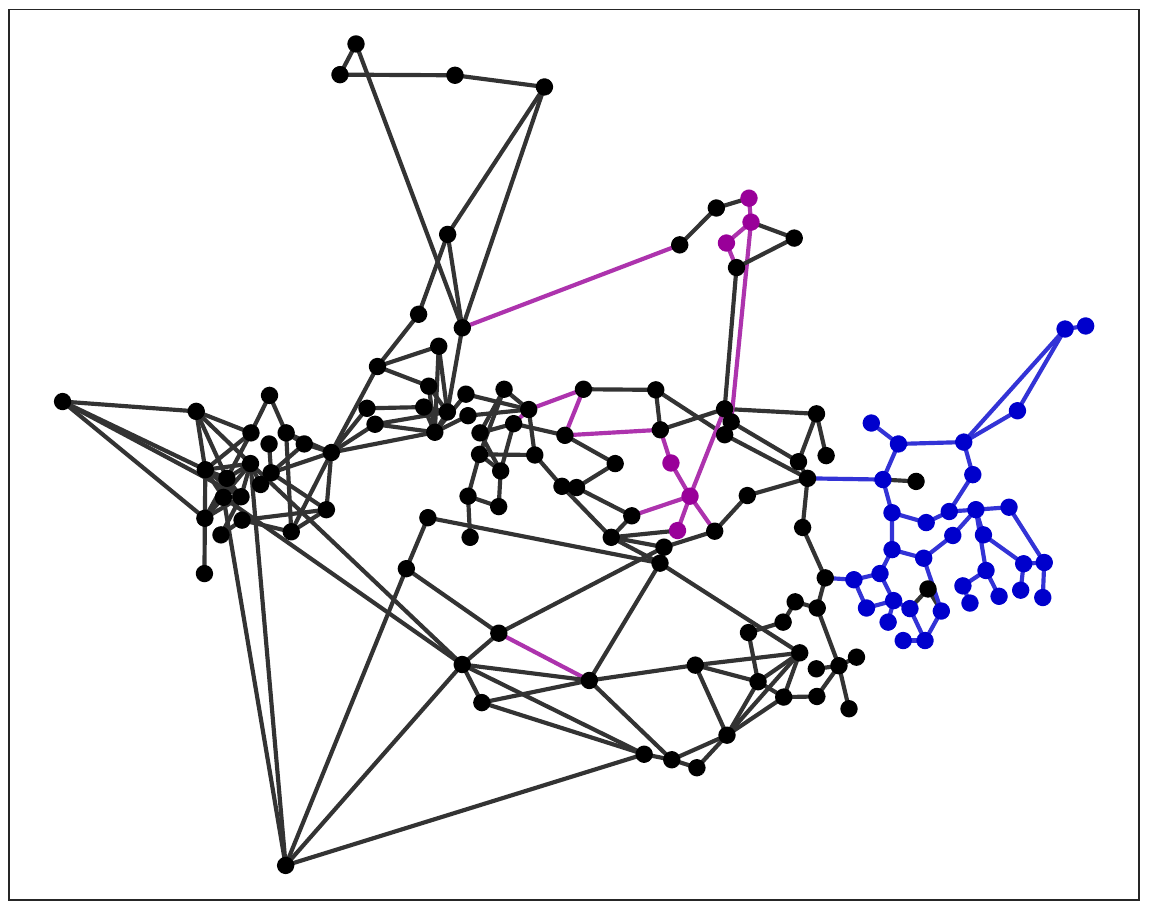}
	\caption{\csentence{Community Structure in CSI}
	Results of Adaptive BRIM community detection on the adjacency matrix of the bipartite instability graph. Small communities (containing fewer than 2 lines or buses) have been subsumed into neighboring clusters.}
	\label{fig:community_res}	
\end{figure}

The Adaptive BRIM algorithm detected 12 distinct communities from the bipartite instability matrix. However, many of these were quite small. In the interest of restricting focus to network-wide swing instabilities, communities containing fewer than 2 buses or lines were reassigned to a neighboring cluster based on a vote of graphical nearest neighbors. This led to a reduced population of 3 dominant communities, plotted in Fig. \ref{fig:community_res}.

The blue community, which corresponds geographically to New England, is a contiguous and highly interconnected nodal group with few connections to the rest of the network. A simple topological community detection scheme applied directly to the NPCC-140 graph would almost certainly have identified it as a highly modular cluster. The purple group, however, is thoroughly embedded within the larger black nodal population. This is a testament to the importance of our dynamics-based approach to community identification: modular graphical structure does not necessarily equate to dynamical coherency.

\section*{The Sensitivity of Network Connections}
\label{sec:sensitive}

Given the diversity of dynamics observed for disruptions of the network, our analysis aims to understand the sensitivity of each node in the power grid to fault tolerances.  By varying the fault severity, we can rank the nodes and their susceptibility to CSI.

\subsection*{Varying Fault Severity}
To identify the lines where a system fault is most likely to generate CSI, we measure responses to faults of varying intensity. Faults in PST simulations are parameterized by two time durations: $\tau_1$ from the application of the fault to the clearing of the near end, and $\tau_2$ from the clearing of the near end to that of the remote end. Generally speaking, a longer fault duration drives the system farther from its initial steady-state configuration, increasing the likelihood of instability.  Indeed, the parametrization of faulty intensity through the $(\tau_1,\tau_2)$ parameter space allows us to characterize the robustness of each node.

\subsection*{Ranking Lines by Sensitivity}
\label{sec:worst_lines}
Working in $(\tau_1,\tau_2)$ parameter space, we identify a domain which captures the onset of instability for most fault locations.  This gives a range of values broad enough so that the lowest values of $\tau_1$, $\tau_2$ yield fully stable dynamics, while the highest values of $\tau_1$, $\tau_2$ lead to instability at many fault locations. The precise choice of bounds is somewhat arbitrary as long as they meet these criteria; the performance over the specified ($\tau_1$,$\tau_2$) region will only ever be used as a relative metric for comparing lines. For each fault location, we repeat simulations over a grid in this domain of parameter space. The performance of each run is quantified by determining the number of buses which go unstable.

\begin{figure}[h!]
	\includegraphics[width=0.9\columnwidth]{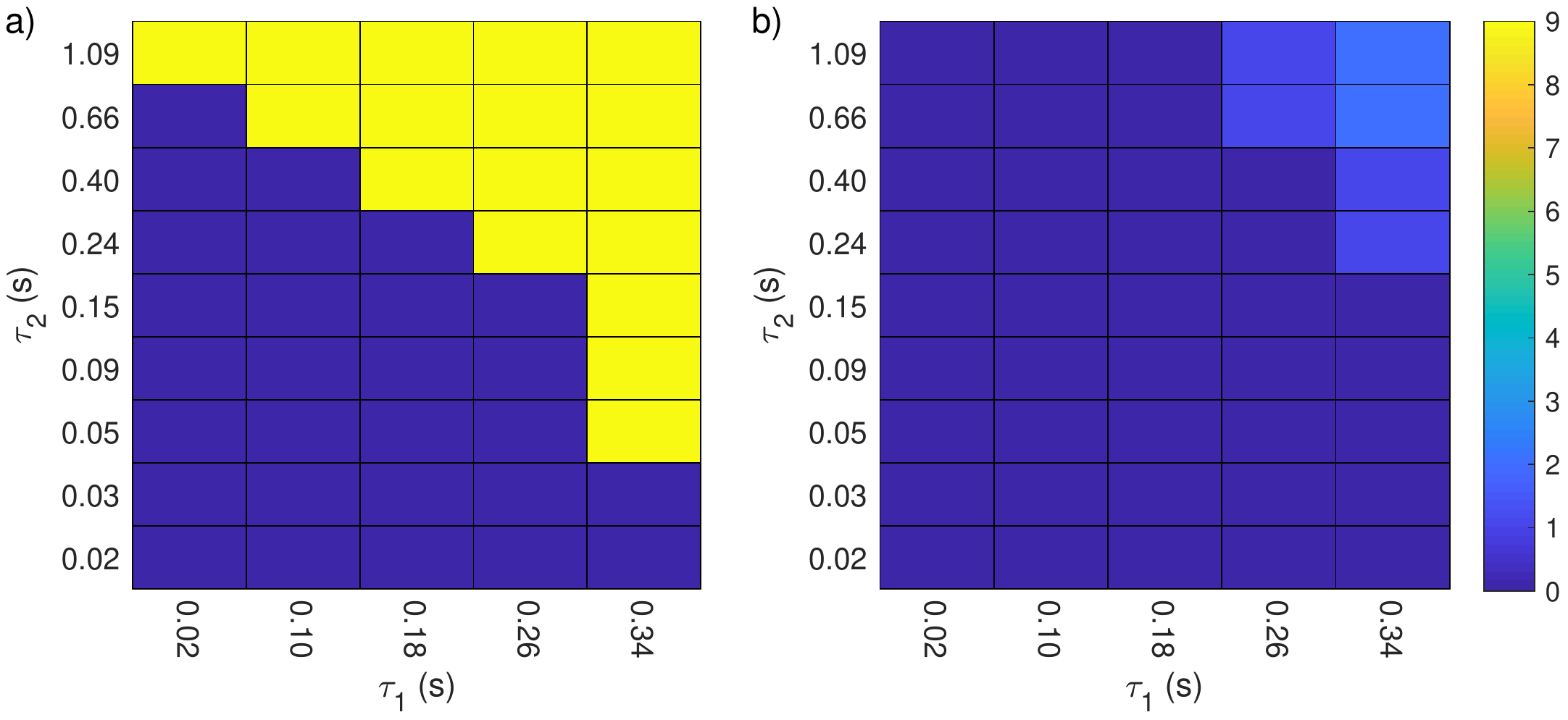}
	\caption{\csentence{Sample Instability Frontiers in Fault-Time Parameter Space}
	Sample results of simulations over a region of fault-time parameter space. Each box represents a simulation with fault times $(\tau_1,\tau_2)$, with its color denoting the number of generator buses which exhibited unstable dynamics. A set of simulations like this was carried out for each line in the network. Examples are presented for a) a fault location likely to induce CSI, and b) a fault location unlikely to induce CSI. Note: the color scale is produced by ennumerating unstable \textit{generator} buses (just those marked by squares in Fig. \ref{fig:NPCC_map}) so a count of 9 corresponds to the whole New England subnetwork.}
	\label{fig:sample_instability_frontiers}
\end{figure}

\begin{figure}[h!]
	\includegraphics[width=0.9\columnwidth]{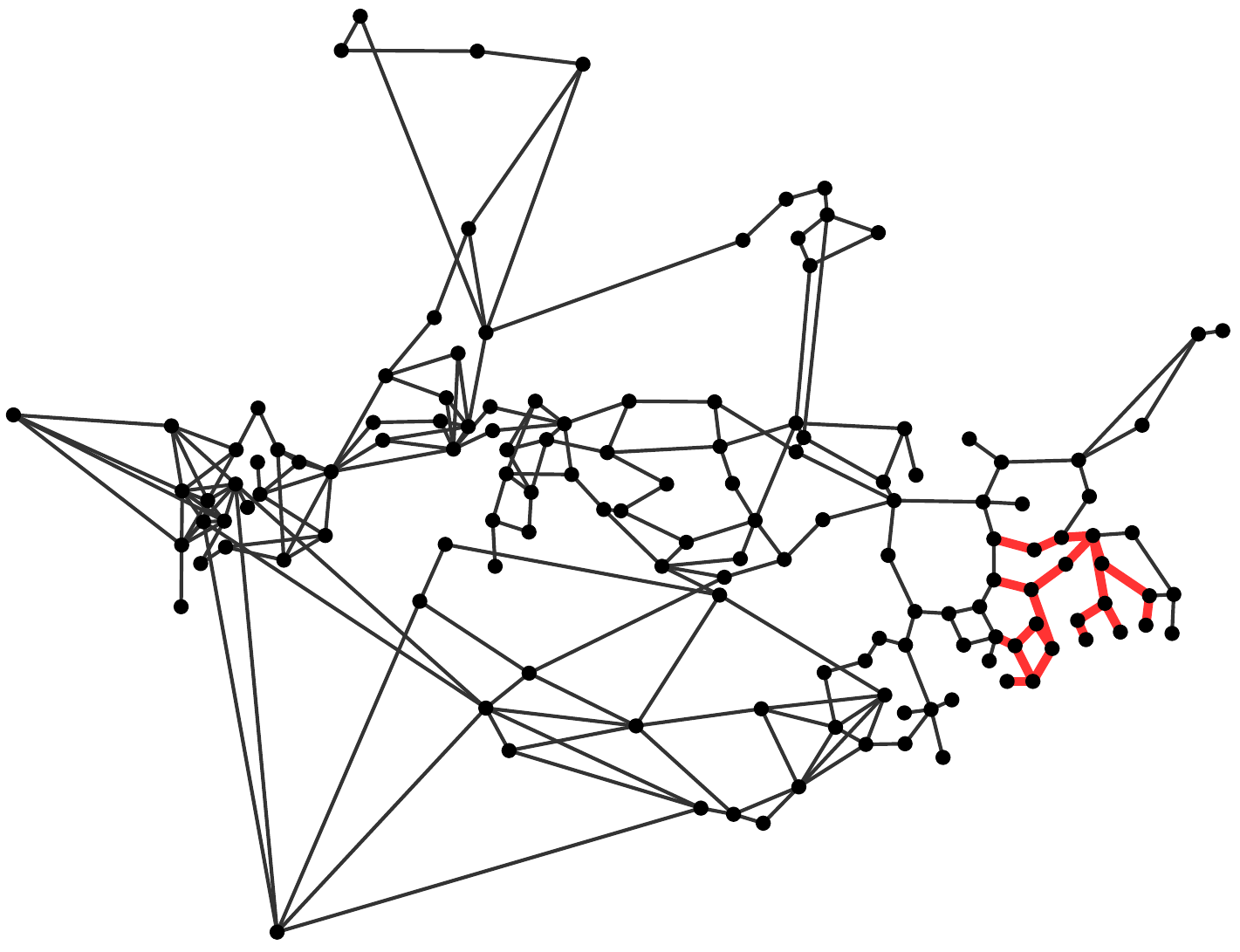}
	\caption{\csentence{Fault Locations Susceptible to CSI}
	Highlighted in red: the twenty most sensitive lines of the network, i.e. those where a fault is most likely to induce CSI}
	\label{fig:worst_lines}	
\end{figure}

The result of this analysis gives an ``instability frontier'' in the space of $(\tau_1,\tau_2)$. As visualized in Fig. \ref{fig:sample_instability_frontiers}, the more sensitive lines of the network (e.g.  Fig.~\ref{fig:sample_instability_frontiers}a) have a frontier which extends farther down toward the bottom-left corner, whereas more robust lines (e.g. Fig.~\ref{fig:sample_instability_frontiers}b) are fully stable until comparatively high values of $(\tau_1,\tau_2)$. In the former case, we observe that a small perturbation to the fault parameters can lead to a big jump in the number of unstable generators. This foreshadows the crucial role of community structure in understanding this behavior: instability often occurs collectively in coherent subnetworks.

Having performed exhaustive simulations for all possible fault locations, we rank lines according to sensitivity by averaging over the unstable nodes obtained at all $(\tau_1,\tau_2)$ parameter combinations. The numerical values obtained are of course sensitive to our choice of the fault-time domain, but they serve as a functional metric of comparison between different lines tested on this domain. Figure~ \ref{fig:worst_lines} illustrates the results of this process, with the 20 most sensitive lines highlighted in red.

These are the lines whose performance we seek to improve through the remainder of this study. The network modifications we consider in the next section belong to a combinatorially large space, so it is necessary to restrict the scope of our analysis wherever possible to avoid having to do a prohibitively large number of simulations. Restricting simulations to these ``worst offender'' fault locations allows us to constrain the parameter space while still treating the cases which are of greatest practical concern with respect to grid stability.

%	MODIFYING NETWORK STRUCTURE
\section*{Engineering Network Structure to Reduce CSI}
\label{sec:eng}
Although the specific criteria which lead a subgroup of buses to coherently desynchronize in a given network are not generally well understood, it is clear that the observed phenomenon of community structure is intimately related to the connectivity configuration of the network. Grids are prone to CSI when they contain a subnetwork which is relatively weakly coupled to its surrounding nodes.  Thus a naive approach to engineering network stability would be to simply add connections between nodes inside and outside this instability prone community. The results we present in this section not only support this intuition, but also show that not all inter-community line additions yield significant improvements to stability. As such, the full simulation-based approach implemented here is necessary to determine \textit{which} inter-community connections contribute the most to the grid's structural robustness.

\begin{figure}[h!]
	\includegraphics[width=0.9\columnwidth]{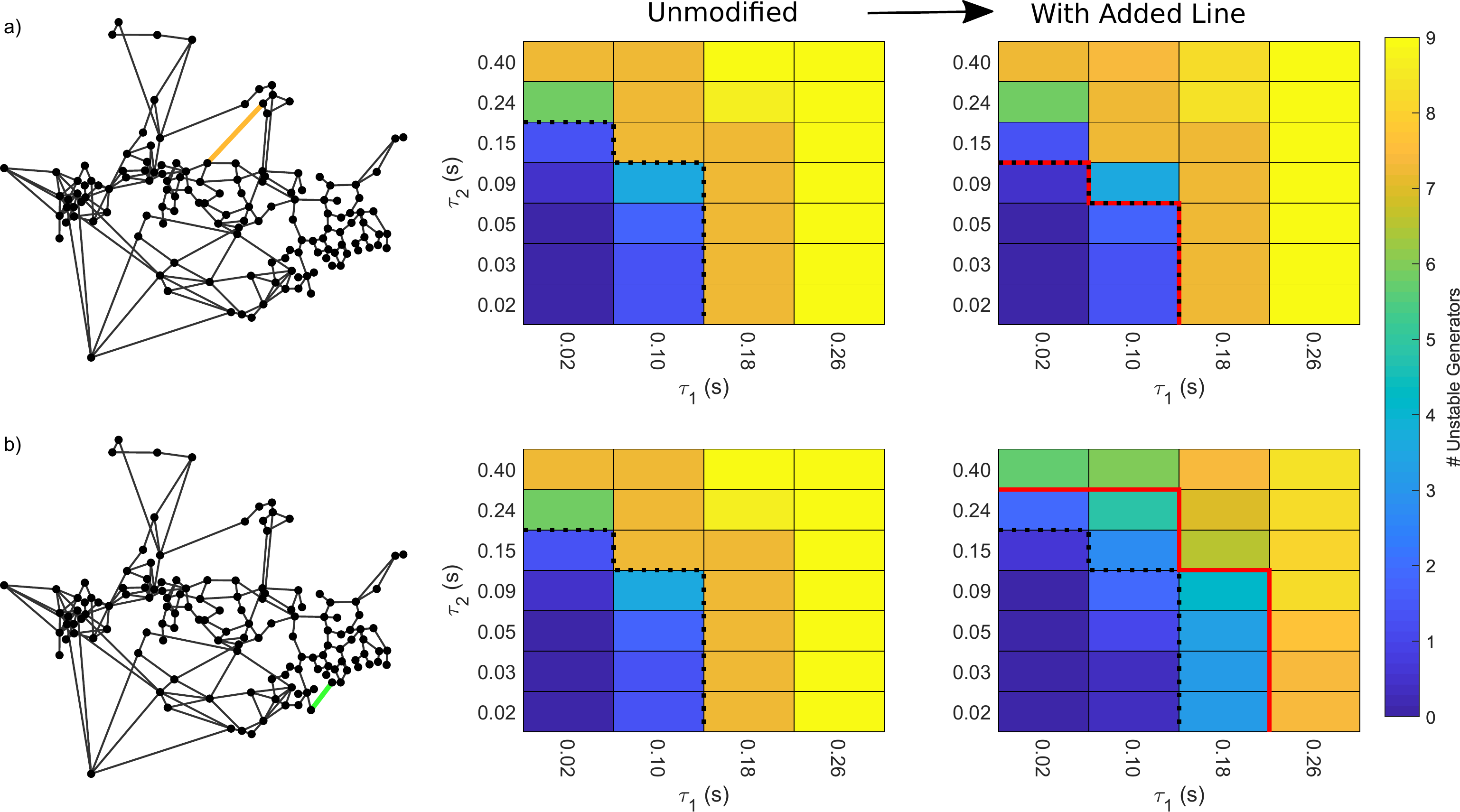}
	\caption{\csentence{Instability Frontiers of Modified Networks}
	For each line addition considered (connections highlighted in color on the left), the mean instability frontier of the unmodified network (center, dotted black line) was compared to that of the network with the new connection (right, solid red line). Some candidates afforded little to no improvement ((a), with added line in orange), while others robustified the network so that it was resilient to considerably more severe faults ((b), with added line in green).}
	\label{fig:instability_frontier_change_annotated}
\end{figure}

\subsection*{Network Modification Protocol}
The approach for assessing how the addition of a transmission line affects the incidence of CSI is as follows:
\begin{enumerate}
\item
A line connecting the two chosen buses (with resistance and reactance specifications taken to be the median of those of the existing lines) is inserted into the PST network specifications.
\item
For each of the most sensitive lines in the unmodified network (i.e. those highlighted in Fig. \ref{fig:worst_lines}), simulations are carried out for all $(\tau_1,\tau_2)$ fault-time combinations in the domain identified previously to obtain an instability frontier (denoted in Fig. \ref{fig:instability_frontier_change_annotated} by a black dotted line). 
\item
Overall performance for each fault location is again obtained by averaging over the sum of unstable generators. These values are then averaged over all tested fault locations to obtain a single plot of the modified network's susceptibility to CSI (shown in the right column of Fig. \ref{fig:instability_frontier_change_annotated}).
\item
The result of this process is compared to that of the unmodified network to obtain a ratio measuring the stability improvement afforded by the added connection.
\end{enumerate}

The set of possible single-line additions to the network is combinatorially large ($N_{\rm{additions}} = N_{\rm{bus}}\left(N_{\rm{bus}}-1\right) - N_{\rm{line}} = 19367$, in the case of the NPCC-140 system). The number of simulations necessary to carry out the above steps for a single network modification is such that it is not computationally feasible to test all possible cases.

We present results on a semi-randomly selected subset of these possibilities (``semi-random'' because they were chosen with a penalty on geographic distance between the buses to be connected, so as to better conform to practical engineering considerations). These candidates are separated into two groups: those which run between two distinct communities, and those which are internal to a single community. 

\subsection*{Network Modification Results}
\begin{figure}[h!]
	\includegraphics[width=0.9\columnwidth]{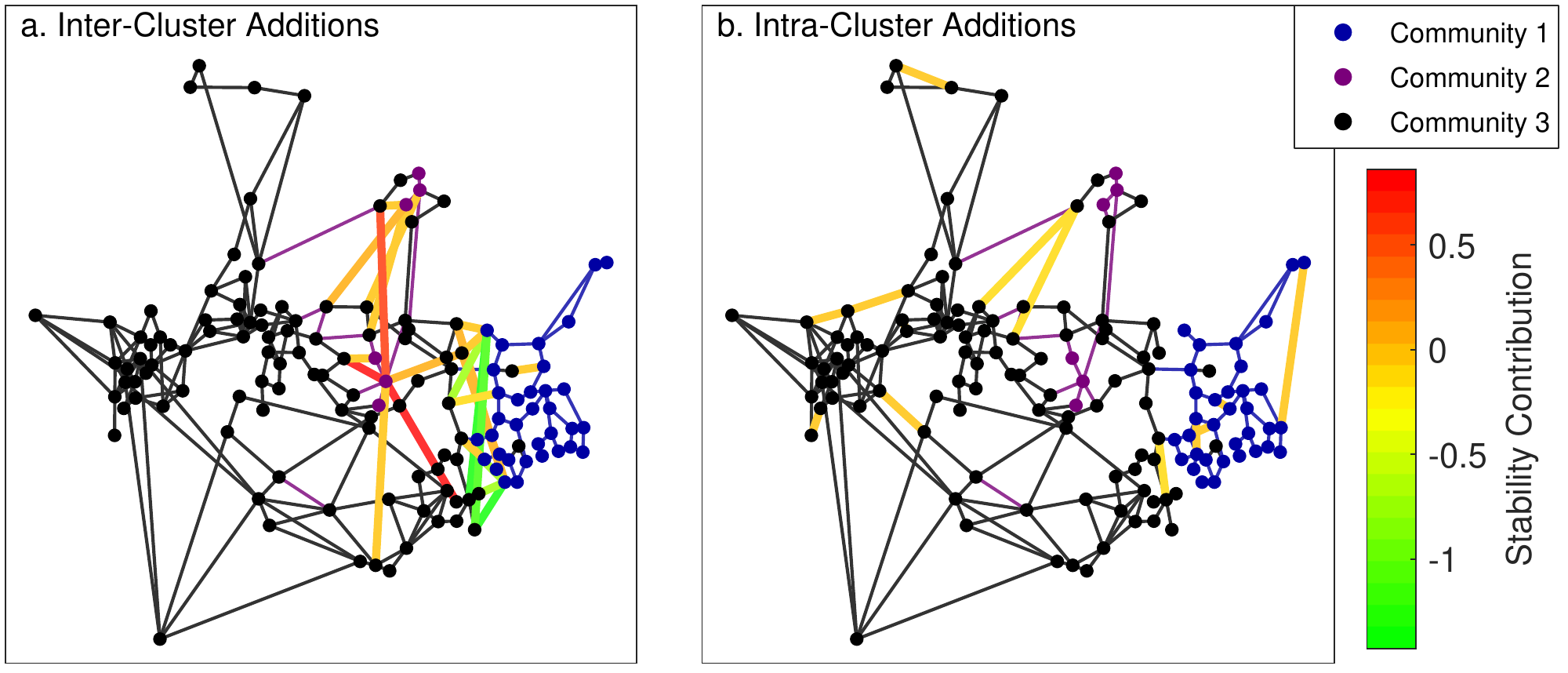}
	\caption{\csentence{Line Additions Colored by Contribution to Stability}
	The NPCC 140-bus system with colored lines representing candidates for additions to the network. Their coloring denotes the extent to which they improved network stability relative to the original network. Green lines afforded the greatest improvement, while red lines left performance largely unchanged. The subnetworks obtained by community detection are colored in black, blue, and purple for reference. Line additions have been separated into a) connections between distinct communities, and b) connections within a single community}
	\label{fig:line_candidate_means_map}	
\end{figure}

\begin{figure}[h!]
	\includegraphics[width=0.9\columnwidth]{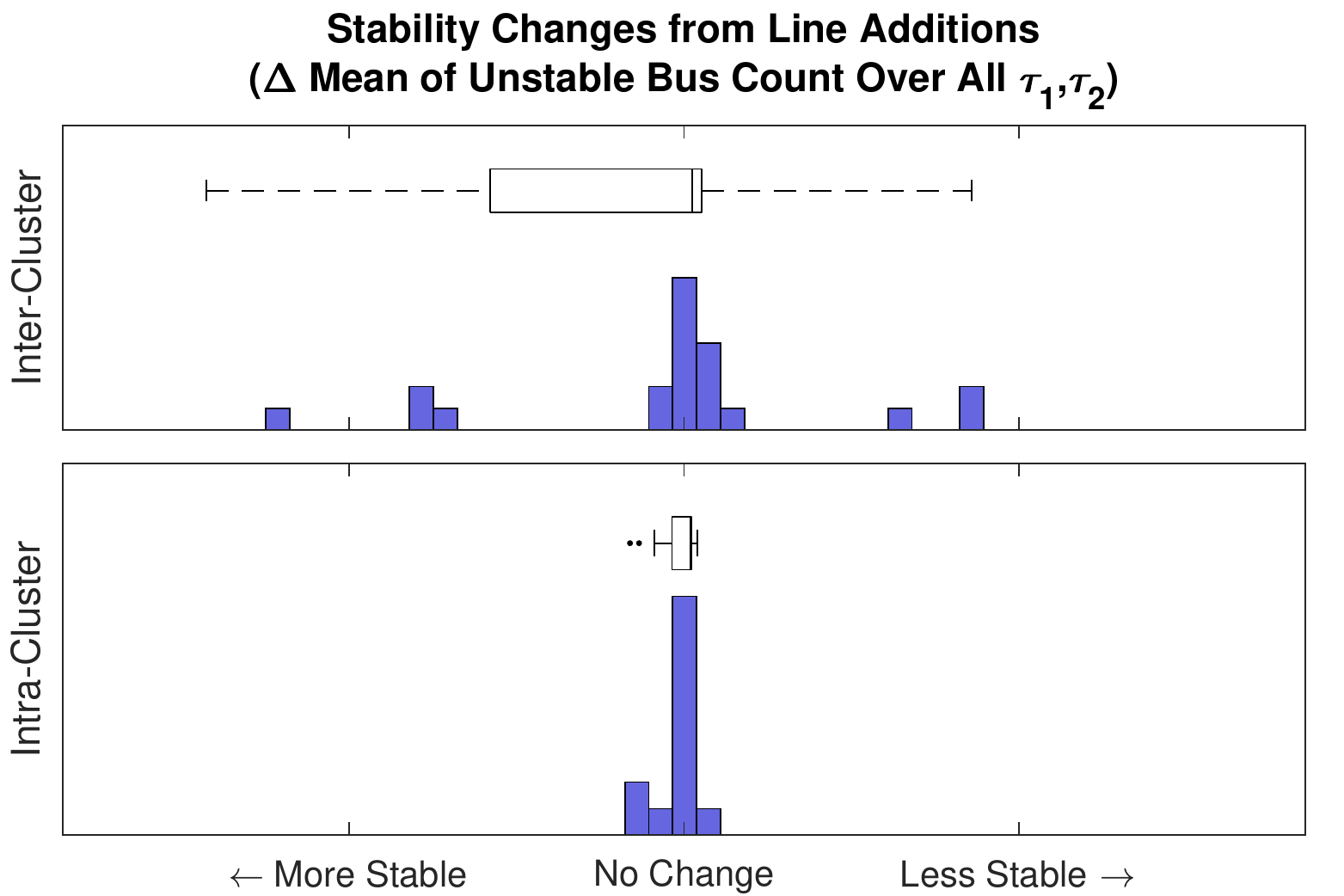}
	\caption{\csentence{Distributions of Stability Change Parameter: Inter- vs. Intra-Community}
	Distributions of stability impact parameter for a) inter-community connections, and b) intra-community connections. From the histogram outliers and the spread denoted by the box plot width, it is clear that the line additions between distinct communities are much more likely to effect significant changes to network stability (positive or negative).}
	\label{fig:line_candidate_means_hist}	
\end{figure}

The candidate lines tested using the above protocol are pictured in Fig. \ref{fig:line_candidate_means_map}, colored according to their performance relative to the original network configuration. Additionally, the distributions of the stability change parameter for each of the two populations of line candidates are presented as histograms and box plots in Fig. \ref{fig:line_candidate_means_hist}.

We observe that every line addition that significantly impacted the network's susceptibility to CSI is one that connects two distinct communities. This suggests that understanding the network's community structure is crucial to identifying candidates for new lines to robustify the grid. The search space of inter-community lines is substantially smaller (
$N_{\rm{additions}} = \frac{1}{2}\sum_i^{N_{\rm{community}}} \left[N_{\rm{bus}}^i \left(N_{\rm{bus}}^{\rm{tot}} -  N_{\rm{bus}}^i \right)\right] - N_{\rm{line}}^{\rm{inter}} = 4191$, in the case of the NPCC-140 system, which represents a 79\% reduction of the search space). Coupled with practical distance considerations (and any other logistical criteria such as local topography or infrastructure), the number of line candidates for which this extensive simulation process must be carried out becomes much more tractable.

Within the inter-community population of line candidates, the criteria for significant robustification are not obvious. Examining the relationship between performance and common topological metrics such as betweenness and degree centrality reveals little to no correlation, which supports the literature consensus \cite{bompard09,bompard10,pagani13,hines10} that these measures are insufficient as predictors of behavior in real power systems. Nonetheless, we have successfully identified a number of lines which do offer a marked improvement even by testing a very small subset of all possible inter-community connections.  This shows that our approach to evaluating robustness of the network can effectively identify potential connections capable of robustifying the power grid network.

\section*{Conclusion}
\label{sec:conclusion}

We propose a computational framework for the analysis of the network level dynamics and stability of power system disturbances.  Our analysis is critical for understanding how the network architecture itself can lead to subgraphs (communities) that are highly susceptible to CSI.  By systematically parametrizing disturbances according to the temporal parameters $\tau_1$ (the time of the fault until it is cleared) and $\tau_2$ (from the clearing time of the near end to that of the remote end), we can assess the effect of each node on the overall stability of the power grid network.

We specifically develop evaluation metrics to (i) identify dynamics-based community substructures in the power grid network, (ii) determine weak points in the network that are particularly sensitive to faults, and (iii) produce an engineering approach for the addition of transmission lines to maximally reduce the incidence of CSI.
For our example of the Northeast power grid, we identify a strong dependence of line sensitivity on the New England subnetwork.  We show how modifying the network's connectivity structure can robustify the network to CSI.  The space of possible connectivity changes is combinatorially large, so we restrict modifications to a tractably small subset of single-line insertions to the original network. We show that community detection can be used to substantially reduce this search space, as line additions which connect separate communities tend to much more significantly impact global stability. We find that the addition of a line can markedly improve the grid's resilience to CSI, but the success is highly dependent on location within the network.

Our analysis provides a versatile diagnostic for the efficacy of adding a particular line to a power grid which is known to be prone to CSI. This is a particularly relevant problem in large-scale power systems, where improving stability by increasing overall network connectivity is not feasible due to financial and infrastructural constraints. Our approach focuses principally on network topology, so its results are fairly robust to small variations in the model parameters used in simulation. This makes it a strong candidate for use in analyzing real-world power systems, as connectivity structure is a characteristic which can be perfectly reproduced in the translation from physical system to simulation model.

\section*{Declarations}
\subsection*{Availability of Data and Material}
All data analyzed in this study was generated using the publicly-available Power System Toolbox for Matlab, which can be downloaded by request from \url{http://www.eps.ee.kth.se/personal/vanfretti/pst/Power_System_Toolbox_Webpage/PST.html}. The parameter definitions for the NPCC-140 Bus System are bundled with the toolbox in the file \texttt{pstdat/data48em.m} 
\subsection*{Competing Interests}
The authors have no competing interests to declare.
\subsection*{Funding}
JNK acknowledges support from the Air Force Office of Scientific Research (FA9550-17-1-0329).  This work was also supported by the U.S. Department of Energy (DOE) Office of Science, Office of Advanced Scientific Computing Research (ASCR) as part of the Multifaceted Mathematics for Rare, Extreme Events in Complex Energy and Environment Systems (MACSER) project. A portion of the research described in this paper was conducted under the Laboratory Directed Research and Development Program at Pacific Northwest National Laboratory (PNNL). PNNL is operated by Battelle for the DOE under Contract DE-AC05-76RL01830.   
\subsection*{Authors' Contributions}
DD was lead author and conducted the computational analysis under the supervision of PhD advisor JNK. All authors participated in conceiving and designing the study. All authors read and approved the final manuscript.
%\subsection*{Acknowledgements}
%Not applicable
%\subsection*{Authors' Information}

%%%%%%%%%%%%%%%%%%%%%%%%%%%%%%%%%%%%%%%%%%%%%%
%%                                          %%
%% Backmatter begins here                   %%
%%                                          %%
%%%%%%%%%%%%%%%%%%%%%%%%%%%%%%%%%%%%%%%%%%%%%%

\begin{backmatter}

%\section*{Competing interests}
%  The authors declare that they have no competing interests.
%
%\section*{Author's contributions}
%    Text for this section \ldots
%
%\section*{Acknowledgements}
%  Text for this section \ldots
%%%%%%%%%%%%%%%%%%%%%%%%%%%%%%%%%%%%%%%%%%%%%%%%%%%%%%%%%%%%%
%%                  The Bibliography                       %%
%%                                                         %%
%%  Bmc_mathpys.bst  will be used to                       %%
%%  create a .BBL file for submission.                     %%
%%  After submission of the .TEX file,                     %%
%%  you will be prompted to submit your .BBL file.         %%
%%                                                         %%
%%                                                         %%
%%  Note that the displayed Bibliography will not          %%
%%  necessarily be rendered by Latex exactly as specified  %%
%%  in the online Instructions for Authors.                %%
%%                                                         %%
%%%%%%%%%%%%%%%%%%%%%%%%%%%%%%%%%%%%%%%%%%%%%%%%%%%%%%%%%%%%%

% if your bibliography is in bibtex format, use those commands:
\bibliographystyle{bmc-mathphys} % Style BST file (bmc-mathphys, vancouver, spbasic).
\bibliography{ANS_Paper_Manuscript_v2_ArXiV}      % Bibliography file (usually '*.bib' )

\end{backmatter}
\end{document}